# Magnéli phases doped with Pt for photocatalytic hydrogen evolution

Ewa Wierzbicka[1#], Maximilian Domaschke[2#], Nikita Denisov[1], Dominik Fehn[3], Imgon Hwang[1], Marlena Kaufmann[2], Babette Kunstmann[2], Jochen Schmidt,[2] Karsten Meyer[3], Wolfgang Peukert[2*], Patrik Schmuki[1,4,5*]

1. Department of Materials Science WW-4 LKO, University of Erlangen-Nuremberg Martensstrasse 7, 91058 Erlangen, Germany.
2. Institute of Particle Technology, University of Erlangen-Nuremberg, Cauerstrasse 4, 91058 Erlangen, Germany.
3. Department of Chemistry and Pharmacy, Inorganic Chemistry, Friedrich-Alexander University Erlangen-Nürnberg (FAU), Egerlandstrasse 1, 91058 Erlangen, Germany.
4. Department of Chemistry, King Abdulaziz University, Jeddah, Saudi Arabia.
5. Regional Centre of Advanced Technologies and Materials, Department of Physical Chemistry, Faculty of Science, Palacky University, Slechtitelu 11, 783 71 Olomouc, Czech Republic

# indicates equal contribution of the work by the authors.

**Corresponding Author**

* Patrik Schmuki E-mail: schmuki@ww.uni-erlangen.de,

* Wolfgang Peukert E-mail: wolfgang.peukert@fau.de;

Abstract

Defined substoichiometric titanium oxides ($Ti_xO_{2x-1}$ with $3 \leq x \leq 10$) called Magnéli phases have been investigated mostly for their unusual high conductivity and metal-like behavior. In photocatalysis, Magnéli phase containing titania particles have been reported to provide favorable charge separation resulting in enhanced reaction efficiency. In the current work we describe a one-step synthesis of Magnéli-containing mixed phase nanoparticles that carry directly integrated minute amounts of Pt. Phase optimized nanoparticles that contain only a few hundred ppm Pt are very effective photocatalysts for $H_2$ evolution (they provide a 50-100 times higher $H_2$ evolution than plain anatase loaded with a similar amount of Pt). These photocatalysts are synthesized in a setup combining a hot-wall reactor that is used for $TiO_x$ synthesis with a spark generator producing Pt nanoparticles. Different reactor temperatures result in various phase ratios between anatase and Magnéli phases. The titania nanoparticles ($\approx$ 24 - 53 nm) were characterized using XRD, HRTEM, XPS and EPR spectra as well as ICP-OES analysis. The best photocatalyst prepared at 900 ºC (which consists of mixed phase particles of 32% anatase, 11% rutile and 57% Magnéli phases loaded with 290 ppm of Pt) can provide a photocatalytic $H_2$ evolution rate of $\approx$ 5432 $\mu$mol h$^{-1}$ g$^{-1}$ for UV and $\approx$ 1670 $\mu$mol h$^{-1}$ g$^{-1}$ for AM1.5 illumination. For powders converted to higher amounts of Magnéli phases (1000 ºC and 1100 ºC), a drastic loss of the photocatalytic $H_2$ generation activity is observed. Thus, the high photocatalytic efficiency under best conditions is ascribed to an effective synergy between multi-junctions of Magnéli titania and Pt that enable a much more effective charge separation and reaction than conventional Pt/anatase junctions.

Magnéli phases are a set of substoichiometric titanium oxides – such as $Ti_3O_5$, $Ti_4O_7$, $Ti_5O_9$, with a general chemical formula $Ti_nO_{2n-1}$ where n varies from 3 to 10; these compounds were firstly described in the 1950s by Magnéli et al. [1–4]. The oxygen deficient deviation from the typical $TiO_2$ stoichiometry is regularly introduced throughout the lattice; it results in delocalized electrons in the *d* band which in turn are the origin of the very high electric conductivity of these compounds [5–8]. In fact, conductivities can reach values comparable to carbon or even to some metals [9–12]. However, most Magnéli phases show a semiconductive behavior at low temperatures and a metallic behavior at higher temperatures [13–17]. For example, $Ti_4O_7$ [7] and $Ti_5O_9$ [18] show a semiconductor-to-metal transition at 149 °K and 139 °K, respectively, and their behavior at room temperature therefore is metal-like. $Ti_3O_5$ can be both a metallic conductor in case of γ-$Ti_3O_5$, while δ-$Ti_3O_5$ is a semiconductor (at 150 °K) with a narrow band gap of 0.07 eV [19]. Moreover, theoretical considerations show that e.g., the filled metallic band of $Ti_4O_7$ is close below the conduction band edge of $TiO_2$ [20]. Other work reports bands positions of $Ti_5O_9$ (with a 2 eV band gap) that are lowered relative to the VB and the CB of $TiO_2$, thus photogenerated electrons in junction with $TiO_2$ transfer to the CB level of $Ti_5O_9$, while holes transfer to the VB of $TiO_2$ [21]. On the other hand, the combination of anatase with metallic suboxides enables charge separation in analogy with semiconductor/metal junctions. Therefore, mixed phase Magnéli junctions deliver cascades of electronic states, which makes these materials highly interesting as intrindic co-catalysts or sensitizers for titania in photocatalytic applications. In a previous article we described the beneficial effect of $Ti_4O_7$ on an anatase nanoparticle photocatalyst for $H_2$ production. We showed that this combination allowed for a significant hydrogen production, without the use of any further co-catalyst [22]. In the present work we explore the synergistic effect of Magnéli type multiphase titania with traces of Pt dosed directly during the one-step nanoparticle synthesis in a hot-wall flow reactor. In classic $TiO_2$ based photocatalysis, Pt is the ultimate benchmark in effectiveness as co-catalyst for $H_2$ generation. Pt on $TiO_2$ accelerates most effectively the kinetically hampered photocatalytic $H_2$ evolution reaction as originally found in 1978 [23]. The effect of Pt is twofold: *i*) Pt and the n-type semiconductor $TiO_2$ form a heterojunction (Schottky junction) [24,25] that enables electron trapping in the metal (the metal acts as 'electron sink'), which leads to prolonging electron lifetime and limits charge recombination, *ii*) Pt acts as efficient $H^0$ recombination catalyst ($2H^0 \rightarrow H_2$). Optimized classic titania photocatalysts (anatase or P25) are loaded with Pt particles at a level of 0.2 - 0.5 wt% (2000 – 5000 ppm). The decoration is typically carried out by a secondary chemical treatment after particle synthesis using for instance wet impregnation approaches. In contrast, in the present work we introduce an in-situ incorporation of extremely small amounts of Pt (dosing only a few hundred ppm) directly during continuous flow synthesis of Magnéli phase nanopowders (for reactor and details see **Supporting Information**). This continuous flow reactor [22,26,27] (**Figure S1a**) allows the formation of a wide range of mixed phase Magnéli type or pure Magnéli type nanoparticles ($\leq$ 55 nm) by variation of reaction parameters. The most crucial process parameter of the setup (see **Supporting Information**) is the reactor temperature that, in our case, was varied from 500 to 1100 °C.

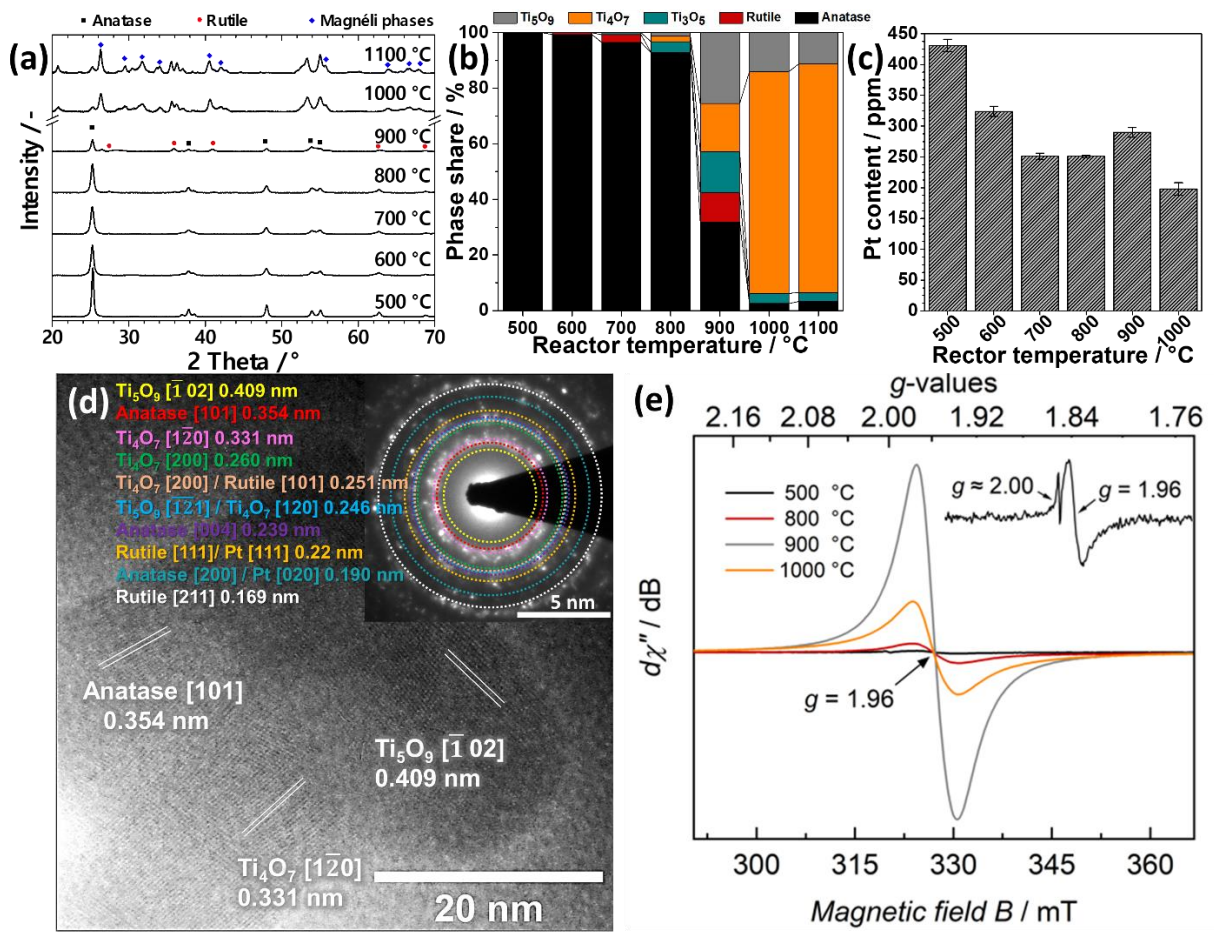

***Figure 1.*** *(a) XRD patterns for the synthesized powders prepared from 500 ºC to 1100 ºC. (b) Phase composition analysis based on Figure 1a. (c) The results of quantification of the Pt content in the titania samples performed by ICP-OES, (d) HRTEM images for sample prepared at 900 ºC. Inset: SAED analysis for the sample prepared at 900 ºC (e) EPR spectra for powders prepared from 500 ºC, 800 ºC, 900 ºC and 1000 ºC, measured at 95 K.*

Under the conditions given in the **Supporting Information**, for reactor temperatures in the range of 500 °C up to ≈ 700 °C, nanoparticles are produced that consist of pure anatase (**Figure 1**). For temperatures > 700°C, various phase transitions occur, namely to rutile and different Magnéli phases, as evident from XRD spectra (**Figure 1a, 1b**). For samples synthesized at different temperatures the phase composition can be obtained from Rietveld refinement of the diffractograms is shown in **Figure 1b**. These data illustrate that at 500 °C pure anatase phase is formed while at higher temperatures (600 ºC and 700 ºC) small amounts of rutile appear (>1 and 3% respectively). At 700 ºC also the presence of Magnéli phases (1%) is detected. With increasing reactor temperature, the content of Magnéli phases becomes significantly higher; at synthesis temperatures of 1000 ºC and higher, Magnéli phases represent the major phases in the nanoparticles. Particularly interesting in the context of photocatalysis are the multi-phase particles between 800 and 1000 °C. For example, the sample generated at 900 ºC contains 32% anatase, 11% rutile and 57% Magnéli phases (with 15% $Ti_3O_5$, 17% $Ti_4O_7$ and 25% $Ti_5O_9$), well in line with PDF data sheets: $Ti_3O_5$ (No. 01-072- 2101, $Ti_3O_5$, Orthorhombic, Cmcm space group), $Ti_4O_7$ (No. 01-072-1724, $Ti_4O_7$, Anorthic, A-1 space group), $Ti_5O_9$ (No. 01-071-0627, $Ti_5O_9$, Anorthic, P-1 space group). It is noteworthy that no Pt peaks can be detected from XRD, nor can particles be detected by SEM, since the concentrations of Pt used are only a few 100 ppm. To confirm the presence of Pt and determine its concentration, inductively coupled plasma optical emission spectroscopy (ICP-OES) was performed. **Figure 1c** gives an overview of the Pt content for the different particles. Please note the incorporated amount is 10-20 times less than the widely used Pt loading on anatase in the range of 0.2 - 0.5 wt%.

SEM images of particles produced at different temperatures are presented **Figure S1 b-h. Figure S2a** gives the median diameters and the geometric standard deviation (GSD) results of particle size distributions for the different powders; additionally the detailed diameter distributions of 500 ºC and 900 ºC samples are shown in **Figure S2b**. At temperatures of 1000 °C and higher, sintering becomes important and leads to an increase in particle size. Overall, the average particle size increases from approximately 24 nm at 800 ºC to 53 nm at 1100 ºC. For temperatures of 1000 ºC and above, clearly the formation of $Ti_4O_7$ becomes dominant and only < 3% anatase is present.

The presence of Magnéli phases in the 900 ºC sample can also be detected by HRTEM imaging based on lattice spacing and using SAED analysis (**Figure 1d and S3**). Based on the observed lattice fringes, it is possible to confirm presence of anatase [101] with a d spacing of 0.354 nm and [004] with 0.239 nm, rutile [111] 0.221 nm, $Ti_4O_7$ [1$\bar{2}$0] 0.331 nm and $Ti_5O_9$ [102] 0.409 nm. It is worth noticing that **Figure 1d** shows a particle with a multi-junction between the Magnéli phases $Ti_4O_7$ and $Ti_5O_9$ with anatase. In spite of extended screening (more than 50 different position checked), no Pt particles could be identified by HRTEM imaging or from SAED, which is not only related to its very small concentration (290 ppm for the sample synthesized at 900 ºC) but also indicates that the Pt deposits are present in a very small size and with an "unlikely to detect" distribution.

XPS measurements for the samples prepared at 500 ºC, 900 ºC and 1000 ºC are shown in **Figure S4**. The XPS data after deconvolution of the O1s peak (**Figure S4a**) show three contributing peaks. The largest peak at 530.1 eV can be attributed to lattice oxygen in the anatase $TiO_2$ position. Its intensity decreases with increasing synthesizing temperature (from 88.78% at 500 ºC to 53.26% at 1000 ºC) while the peak at 532.3 eV that corresponds to "defective" oxygen species typical for Magnéli phases [22,28–30] increases. The peak area increases from 30.17% at 900 ºC to 33.97% at 1000 ºC. The peak at the highest binding energy of 533.9 eV is typically related to water adsorbed on the photocatalyst surface [31]. **Figure S4b** presents the Ti2p peak for the sample prepared at 500 °C, which shows the characteristic for $TiO_2$ doublet located at 458.7 and 464.5 eV. In case of powders fabricated at 900°C and 1000 °C, additional signals appears from reduced forms of $TiO_2$, most significantly a peak doublet that can be assigned to $Ti^{3+}$ (457.2 and 462.9 eV). For Pt no significant peak could be detected as the amount is below XPS detection limit (**Figure S4c**). Thus, a reliable overall quantitative Pt determination was only possible by ICP-OES, as described above (**Figure 1c**).

Electron paramagnetic resonance (EPR) spectroscopic measurements were carried out with all investigated samples (from 500 °C to 1000 °C) as shown in **Figure 1e**. In the spectrum of the anatase sample (prepared at 500 °C), only trace amounts of $Ti^{3+}$ sites, $S = ½$ in a tetragonal field ($D_{4h}$) of $TiO_2$ lattice, are detected, centered at $g \approx 2.00$ **(Figure 1e inset)** [32]. Additionally a very low-intensity, broad EPR signal is present at $g_{iso} = 1.96$ as determined by EPR simulation **(Figure S5)**. Since Magnéli phases show a significant deviation from the $TiO_2$ stoichiometry [29], oxygen vacancies and $Ti^{3+}$ states are regularly present in the crystal structure of the samples Therefore, samples synthesized at higher temperatures (800 °C, 900 °C and 1000 °C), where XRD indicates the presence of Magnéli phases, the signals at $g_{iso} = 1.96$ are significantly more defined and more intense. According to the literature, this signal is attributed to a $(Ti^{3+}\text{-}V_o\text{-}Ti^{4+})^{+1}$ center, which consists of an electron localized at an oxygen vacancy with a possible charge transfer between the two Ti sites [33]. This signal at $g_{iso} = 1.96$ increases in intensity as the synthesis temperature increases, this effect is observed until the synthesis temperature reached 900 ºC. For samples synthesized at 1000 ºC the intensity of paramagnetic resonance is dropping again. This may be explained by the fact that even though the sample synthesized at 1000 ºC contains a larger share of Magnéli phases, the sample obtained at 900 ºC contains the highest amount of the $Ti_3O_5$ phase which shows the largest sub-stoichiometry among all Magnéli phases (in terms of density of oxygen defects compared to $TiO_2$), and therefore provides the highest concentration of paramagnetic centers.

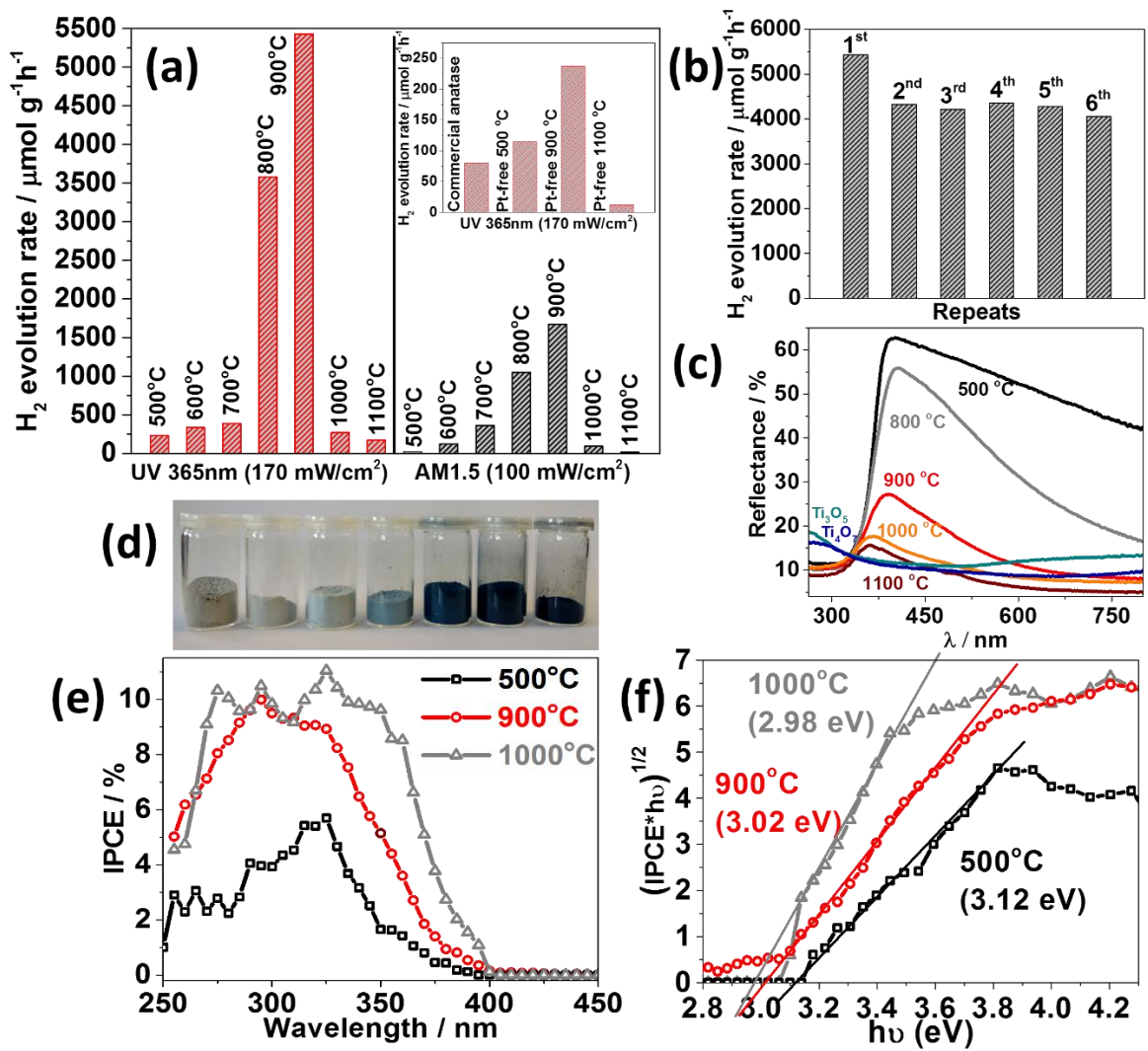

*Figure 2.* (a) Photocatalytic H$_2$ evolution from powders synthesized at different temperatures between 500 ºC and 1100 ºC, under 365 nm LED (170 mW cm$^{−2}$) or AM1.5 (100 mW cm$^{−2}$) solar simulator illumination measured in a 50 vol% methanol–water electrolyte. Inset: experiment conducted for commercial anatase, and powders obtained by the same synthesis method but without Pt. (c) Cycling experiment of H$_2$ evolution under 365 nm LED from the 900 ºC sample that shows the highest photocatalytic activity. (c) Reflectance spectra of the selected powders. (d) Optical images of the samples. (e) IPCE measurements for the as-synthesized materials as electrodes and (f) band gap evaluation from spectra in (e).

**Figure 2a** shows the photocatalytic H$_2$ evolution performance of the single step synthesized powders under AM 1.5 (100 mW cm$^{−2}$) solar simulator or ultraviolet light (λ= 365 nm, 170 mW cm$^{−2}$) illumination. These measurements were carried out using suspensions of the particles in a 50% methanol aqueous solution (experimental details are given in the **Supporting Information**).

For either light source, the photocatalytic $H_2$ evolution efficiency is drastically increasing for synthesis temperatures up to 900 ºC, while it strongly drops for higher synthesis temperatures. Pt functionalized materials showed a significantly higher $H_2$ yield than Pt free powders (**Figure 2a inset**). Remarkably, in comparison with anatase (containing 430 ppm Pt) a significantly higher activity is obtained for powder synthesized at 800 ºC (containing 250 ppm Pt) but the optimum is reached at 900 ºC (290 ppm Pt). As the differences in Pt loading are relatively small and as the anatase powder synthesized at 500 °C contains more Pt than the powder synthesized at 900 °C, the huge difference in $H_2$ production must be ascribed to the phase composition of the powders. For this multi-phase powder, photocatalytic $H_2$ evolution rates of $\approx$ 5432 µmol h$^{-1}$ g$^{-1}$ for UV and $\approx$ 1670 µmol h$^{-1}$ g$^{-1}$ for AM1.5 are achieved. This is 50-100 times higher than for the Pt dosed anatase powder. Obviously, this strong enhancement is a result of the combination of anatase with Magnéli phases and their interaction with the very small amount of Pt. Obviously, still a reasonable amount of anatase needs to be present in the catalyst to obtain this strong catalytic enhancement. The latter is evident from the fact that a further increase of the share of Magnéli-phases in the powders (> 1000 ºC) leads to diminished $H_2$ evolution. At the same time, Pt- free samples show less activity even when compared with the strongest hydrogen evolution for the powder synthesized at 900 ºC (237 µmol h$^{-1}$ g$^{-1}$). Photocatalytic cycling experiments (**Figure 2b**) show that the $H_2$ generation activity is only slightly dropping after first cycle and then is stable over the following intervals.

Reflectance spectra (**Figure 2c**) of the powders depicted in **Figure 2d** show for the 500 ºC sample the typical anatase behaviour where only light below the band-gap of anatase (3.2 eV $\approx$ 350 nm) is absorbed. In contrast, for 1000 ºC and 1100 ºC where samples consist almost entirely of Magnéli phases, a drastic drop of reflection in the visible light range is observed. This is in line with reflectance measurements performed with untreated commercial Magnéli $Ti_3O_5$ and $Ti_4O_7$ powders that show typical metallic features. Partially converted samples (800 °C and 900 °C) show an increasing absorption in the visible light (as frequently discussed in literature for grey to black titania [34,35]). Samples of optimal composition for photocatalysis (900 °C) provide a combination of those features, showing a band gap related to $TiO_2$, with the typical grey tail into the visible range.

In stark strong contrast to these absorption measurements are the characteristics obtained from photocurrents spectra **Figure 2e,f**. While for the plain anatase the expected distinct onset of the photocurrent at energies of 3.12 eV corresponding to the bandgap is observed, for the powder produced at higher temperatures only a slight lowering of the band gap to 3.02 eV and 2.98 eV is observed [22,30,35,36]. That is, no visible light response can be detected for any of the samples. This difference to the light absorption measurements shows that absorbed light in the visible range does not provide sufficiently mobile excited charge carriers (electrons / holes) to be able to contribute considerably to a photocurrent. This is in line with the observation that visible light generated charge carriers do not lead to measurable $H_2$ generation.

From a semi-quantitative assessment of the solid state electric conductivity of the different powders (**Figure 3a,b**; see the **Supporting Information** for details) one can see a drastic increase in the observed conductivity with higher synthesis temperature and an increased amount of Magnéli phase in the powders. The measurements for reference $Ti_3O_5$ and $Ti_4O_7$ powders show full metallic behavior (resistances in the range of several Ω and an entirely linear behavior in the 2-point measurements is obtained). All powders synthesized in our reactor that contain some amount of anatase exhibit the behavior characteristic for a semiconductor junction (a non-linear I-V curve) and resistance values in the MΩ range. This finding is therefore fully in line with the reflectance measurements indicating $Ti_3O_5$ and $Ti_4O_7$ to be metallic at RT. The observed conductivity increase allows to conclude that introduction of these metallic suboxides are correlated with the observed $H_2$ generation increase. In agreement with theoretical work on Magnéli phases, one may thus depict the energetic situation of an anatase/ Magnéli phase /Pt junction as outlined in **Figure 3c**. From our results and above considerations one can deduce that in this multi-junction system, anatase is responsible for light absorption and charge carrier formation whereas the metallic Magnéli phases in intimate vicinity aid charge separation. Photogenerated electrons, due to close placement of the metallic band of the Magnéli phases to conduction band of anatase, may extract electrons and may directly aid the reaction with liquid phase or act as a relays to the transport of the electrons to the ultrafine Pt nanoparticles. Since Magnéli-type oxides are characterized by metallic behavior with high electronic conductivity, not only carrier separation but also mobility can be significantly improved. While Magnéli phases can directly mediate electron transfer to the solution, a cascade to Pt seems to be even more efficient. Clearly, the combination of various types of Magnéli phases, ultrafine Pt domains and anatase in a single particle allows for the creation of divers electronic pathways to fast and effectively transport deliver electrons for the $H_2$ generation reaction.

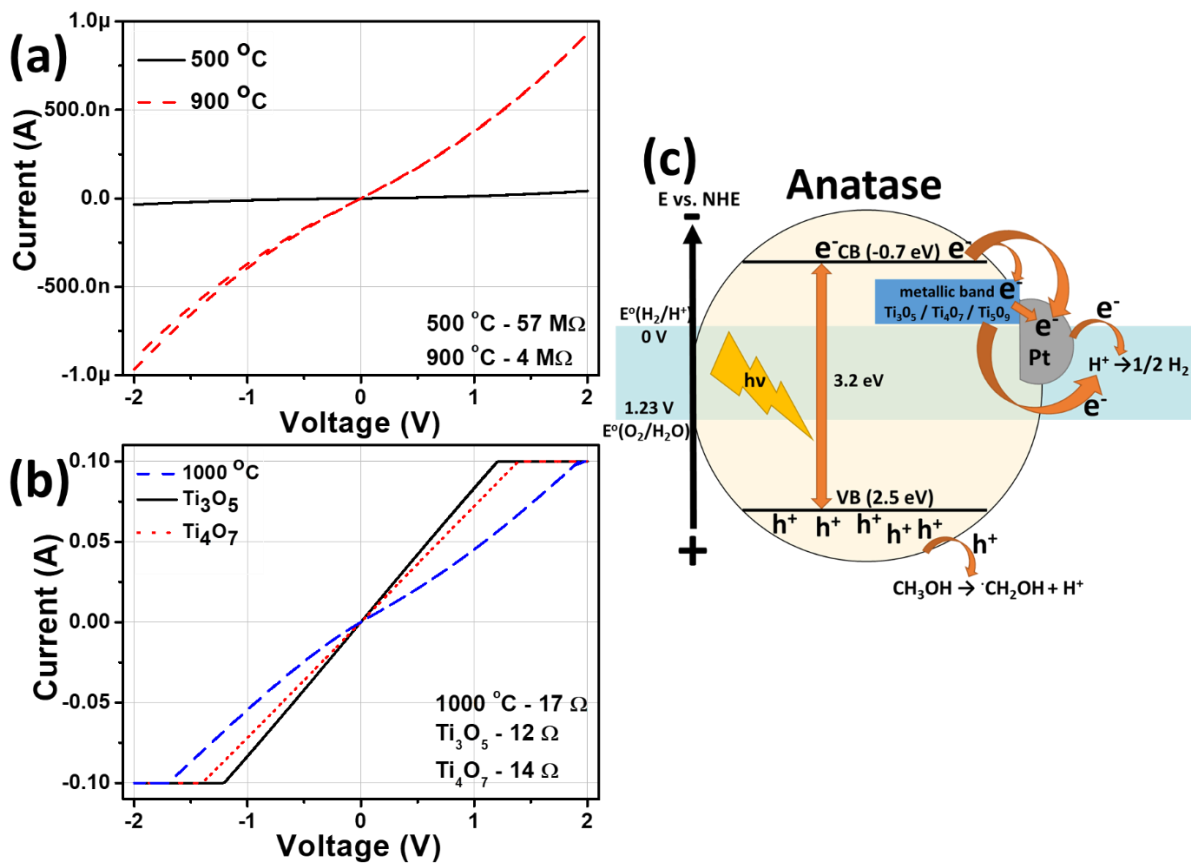

*Figure 3. (a) Solid-state conductivity measurements for samples prepared at 500 ºC, 900 ºC and (b) 1000 ºC, commercial pure $Ti_3O_5$ and $Ti_4O_7$ powders. (c) Schematic drawing for the band alignment between $TiO_2$, Magnéli phase and Pt based on theory.*

This work shows that single nanoparticles, which consist of nanosized anatase / mixed phase Magnéli /Pt junctions within a single particle, are extremely beneficial for photocatalytic hydrogen generation. In our work the photocatalyst is simply formed by a one-step synthesis in a setup combining a hot-wall reactor with a spark generator. The formed nanoparticles contain junctions that provide unique energy states that are able to very efficiently separate and deliver charge carriers to the interphase between catalyst and solution as well as provide a rapid $H_2$ formation reaction. The anatase phase is mostly responsible for charge carrier formation while the mixed Magnéli phases with metallic bands provides a high electronic conductivity and a cascade of electronic states that ensures a high carrier separation and mobility through the catalyst. It is possible that the latter also takes part in the direct transfer of electrons to the solution, however main contribution to the high efficiency of this process is provided by Pt doping. In such a multiphase particle, using an extremely small amount of noble metal is much more effective than in a plain anatase particle. Among the investigated powders, the optimal mixed phase particles consist of 32% anatase, 11% rutile and 57% Magnéli phases (about 15% $Ti_3O_5$, 17% $Ti_4O_7$ and 25% $Ti_5O_9$). These nanoparticles show the highest photocatalytic $H_2$ evolution rate (≈ 5432 μmol $h^{-1}$ $g^{-1}$ for UV and ≈ 1670 μmol $h^{-1}$ $g^{-1}$ for AM1.5) which is 50-100 times more efficient than using anatase with a similar Pt loading.

ASSOCIATED CONTENT

**Supporting Information**: Experimental section, drawing of the reactor setup, SEM images of powders, particles size distributions, HRTEM images for 900 ºC sample, XPS spectra, 900 ºC EPR spectrum and its simulation.

AUTHOR INFORMATION

* Patrik Schmuki E-mail: schmuki@ww.uni-erlangen.de,
* Wolfgang Peukert E-mail: wolfgang.peukert@fau.de;

**Notes**

"The authors declare no competing financial interest."

ACKNOWLEDGMENT

The authors would like to acknowledge ERC, German Research Council (DFG) within the framework of its Excellence Initiative for the Cluster of Excellence ''Engineering of Advanced Materials" (www.eam. fau.de) and Development and Education-European Regional Development Fund, project no. CZ.02.1.01/0.0/0.0/15_003/0000416 of the Ministry of Education, Youth and Sports of the Czech Republic for financial support, as well the Center for Nanoanalysis and Electron Microscopy (CENEM).

Title image:

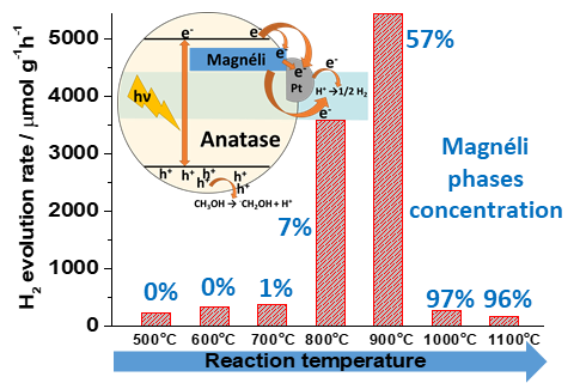

Magneli phase in anatase shows high synergistic interaction with for photocatalytic $H_2$ generation